\documentclass[aps,a4paper,11pt, nofootinbib]{revtex4}
\usepackage{bm}
\usepackage{mathrsfs}
\usepackage{xcolor,color,graphicx,graphics}
\usepackage[all]{xy}
\usepackage{epsfig,subfigure}
\usepackage{latexsym,amssymb,amsmath,amsfonts} 
\usepackage[english]{babel} 
\usepackage[OT1]{fontenc}
\usepackage[latin1]{inputenc}
\usepackage{hyperref}
\usepackage{color,graphicx,graphics,wrapfig,epsf}

\newcommand{\be}{\begin{equation}} 				\newcommand{\ee}{\end{equation}}
\newcommand{\ben}{\begin{eqnarray}} 			\newcommand{\een}{\end{eqnarray}}
\newcommand{\ba}[1]{\begin{array}{#1}} 		\newcommand{\ea}{\end{array}}
\newcommand{\bea}{\begin{equation}\begin{array}{rcl}}
\newcommand{\eea}{\end{array}\end{equation}}

\begin{document}
\title{Generating solutions of Ricci-Based Gravity theories from General Relativity}
\author{Emanuele Orazi} \email{orazi.emanuele@gmail.com}
\affiliation{ International Institute of Physics, Federal University of Rio Grande do Norte,
Campus Universit\'ario-Lagoa Nova, Natal-RN 59078-970, Brazil}
\affiliation{Escola de Ciencia e Tecnologia, Universidade Federal do Rio Grande do Norte, Caixa Postal 1524, Natal-RN 59078-970, Brazil}
%%%%%%%%%%%%%%%%%%%%%%%%%%%%%%%%%%%%%%%%%%%%%%%%%%%%%%%%%%%%%%%%%
\date{today}
%%%%%%%%%%%%%%%%%%%%%%%%%%%%%%%%%%%%%%%%%%%%%%%%%%%%%%%%%%%%%%%%%
\begin{abstract}
%%%%%%%%%%%%%%%%%%%%%%%%%%%%%%%%%%%%%%%%%%%%%%%%%%%%%%%%%%%%%%%%%
We generalize the algorithm that establishes the correspondence between metric-affine Eddington-inspired Born-Infeld (EiBI) gravity and General Relativity (GR) to any bosonic matter sector. Along the way, a polished version of the proof of existence of a metric compatible frame associated to any metric-affine Ricci-Based Gravity (RBG) is presented. Particular attention is given to the problem of generating solutions of a RBG from the GR conterpart, providing a general recipe for the EiBI case. This extends previous results obtained for specific matter that included anisotropic fluids, scalar fields and electromagnetic fields.
%%%%%%%%%%%%%%%%%%%%%%%%%%%%%%%%%%%%%%%%%%%%%%%%%%%%%%%%%%%%%%%%%
\end{abstract}
%%%%%%%%%%%%%%%%%%%%%%%%%%%%%%%%%%%%%%%%%%%%%%%%%%%%%%%%%%%%%%%%%

\maketitle

\section{Introduction}

In the quest for a consistent generalization of General Relativity (GR), the presence of higher order curvature terms in the action is a common feature that arises in low-energy effective description of gravity when quantum correction are included \cite{Donoghue:2017ovt,Vilkovisky:1992pb} and when a renormalization of matter fields in curved space takes place \cite{Birrell:1982ix}. String inspired theories as well point in this direction, suggesting that higher order modifications of GR have to take into account higher-order curvature-related objects \cite{Zwiebach:1985uq,Boulware:1985wk,Mohaupt:2005jd}. However, any extension of GR leads to more involved field equations that require a huge effort to produce observationally relevant exact solutions that could work as discriminators with respect to GR predictions. An interesting proposal that could possibly allows to cast the field equations in a more treatable form is based on the phenomenological equivalence between different theories of gravity at classical level. A stimulating example is provided by the fact that $F(R)$-gravity and Brans-Dicke theory with a scalar potential share the same phenomenology \cite{Flanagan:2003rb,Olmo:2005hc,Sotiriou:2006hs}. Using the metric-affine formalism to describe this correspondence, the Brans-Dicke field turns out to be not propagating in contrast with the formulation of the equivalence in the framework of the metric approach. In fact, this is a signal that keeping some distance from Riemannian geometry some hidden symmetries may arise playing a pivotal role to relate different gravity theories. This is exactly what happens in metric-affine formalism where the affine connection and the metric are independent fields, leading to a symmetry enhancement due to the possibility of transforming metric and connection independently. Particularly interesting is the projective symmetry of the affine connection \cite{Eisenhart} that had driven attention of the scientific community as interesting properties of gravity theories depend on it. For instance, projective transformations has a fundamental role to prove that modified gravity theories based on a simmetric Ricci tensor are ghost free \cite{BeltranJimenez:2019acz}, solving a longstanding problem of higher derivatives gravity for a wide family of theories. This symmetry enhancement is peculiar of minimally coupled theories of gravity described by a Lagrangian that is an arbitrary function of the symmetric Ricci tensor, the so-called Ricci-Based gravities (RBG). 

A fundamental characteristic of RBG is that combining a redefinition of the metric and a specific projective transformation, the field equations associated to the connection can be reduced to a metric compatibility condition \cite{Afonso:2017bxr}. The possibility to recover Riemannian geometry introducing an auxiliary metric, paved the way to propose a correspondence between RBG and GR in presence of scalars and electromagnetic fields, in a series of papers \cite{Afonso:2018bpv,Afonso:2018hyj,Afonso:2018mxn,Delhom:2019zrb}. The main idea consists in a redefinition of matter fields interactions that compensate the deviations from GR. The net effect is that the metric field equations of the RBG are reduced to the Einstein equations associated to a different matter theory without affecting the field content of the original theory. This on-shell equivalence between the space of solutions of different theories is explicitly realized with a mapping between solutions of GR and RBG, putting forward a method to find new solutions on both side of the correspondence. This strategy has been proven to be effective when applied to gravitational theories involving scalar fields \cite{Afonso:2018hyj,Afonso:2019fzv}, electric field and anisotropic fluids \cite{Afonso:2018mxn}. In light of these preliminary results, it is manifest that this procedure has the potential to exploit analytical methods developed withing the framework of GR to explore unknown astrophysical and cosmological scenarios in many modified gravity theories.

In this work, the main steps that led to establish an effective equivalence between RBG and GR will be reviewed in a self-contained way, eventually emphasizing key points to generalize the correspondence to any matter field. In particular, previous results concerning the mapping between solutions in different frames will be extended to any matter sector. Motivated by observational and theoretical appealing property of the EiBI theory \cite{BeltranJimenez:2017doy}, we consider this case as a proper non-trivial example for a detailed and complete description of the mapping procedure, including novel examples when electrodynamics is coupled to gravity. This contribution is intended as a roadmap to generate new solutions for RBG whose physical property will be addressed in future publications.

This work is organized as follows :  in Sec.\eqref{ProjectiveInvariance} we introduce the projective transformation as a reparametrization symmetry of the geodesic equation, discussing the proper tensorial structures that can be included in a gravity action that is invariant under this kind of transformation. In Sec.\eqref{MetricComp} is presented a polished version of the proof that a specific projective transformation and a proper redefinition of the metric can bring any RBG to a Riemannian frame. In Sec.\eqref{Corr-RBG-GR} we review the steps that led to a reduction of the field equations associated to the variation of the metric of a RBG to the Einstein field equations provided that the interactions involving matter fields are properly redefined according to specific differential equations. In Sec.\eqref{GenerSolut} is outlined a general proposal to generate solutions for any RBG gravity and any matter sector, pointing out that the procedure is necessarily model dependent. Sec.\eqref{CorrMattSect} focuses on the resolution of the correspondence-defining equations presented in Sec.\eqref{Corr-RBG-GR} deriving useful parametrizations of the matter Lagrangians. In Sec.\eqref{EiBIsection} it is applied the machinery developed in previous sections to the case of EiBI gravity, discussing general equations that can be used as tools to apply the correspondence to concrete examples. Finally, Sec.\eqref{GenSol} specializes the equations describing the correspondence to the case that involves EiBI gravity and nonlinear electrodynamics showing explicitly the equivalence between gravity theories for specific choices of matter interactions.

%%%%%%%%%%%%%%%%%%%%%%%%%%%%%%%%%%%%%%%%%%%%%%%%%%%%%%%%%%%%%%%%%
\section{Projective Invariance\label{ProjectiveInvariance}}
%%%%%%%%%%%%%%%%%%%%%%%%%%%%%%%%%%%%%%%%%%%%%%%%%%%%%%%%%%%%%%%%%

Assuming the weak equivalence principle as a fundamental principle of a gravity theory, the paths followed by pointlike test particles are solution of the geodesic equation 
\be
\frac{d^2x^\mu}{dt^2} + \Gamma^\mu{}_{\nu\rho}\frac{dx^\nu}{dt}\frac{dx^\rho}{dt} = f(t)\frac{dx^\mu}{dt}\,.\label{GeodEq}
\ee
parametrized in terms of time $t$ in an arbitrary frame. Here, the quantity
\be
f(t)=\Gamma^0{}_{\mu\nu}\frac{dx^\mu}{dt}\frac{dx^\nu}{dt}\label{f(t)}
\ee
has been introduced. Following the standard procedure, the introduction of the affine parameter defined by
\be
\frac{d\tau}{dt} = \xi \, e^{\int{f(t)dt}}\,,
\ee
cancels the ficticious drag force on the right hand side of \eqref{GeodEq}, reducing the geodesic equation to
\be
\frac{d^2x^\mu}{d\tau^2} + \Gamma^\mu{}_{\nu\rho}\frac{dx^\nu}{d\tau}\frac{dx^\rho}{d\tau} = 0\,.\label{SimplGeodEq}
\ee
Taking advantage of the independence of the affine connection with respect to the metric, one can consider a transformation of the form
\be
\hat\Gamma^\mu{}_{\nu\rho} = \Gamma^\mu{}_{\nu\rho} + \gamma^\mu{}_{\nu\rho}\,.\label{ConnTransf}
\ee
that applied to the geodesic equation \eqref{GeodEq} gives
\be
\frac{d^2x^\mu}{dt^2} + \hat\Gamma^\mu{}_{\nu\rho}\frac{dx^\nu}{dt}\frac{dx^\rho}{dt} = \left(f(t)\delta^\mu_{\rho}+ \gamma^\mu{}_{\nu\rho}\frac{dx^\nu}{dt}\right)\frac{dx^\rho}{dt}\,.\label{GenGeodEq}
\ee
Reminding the expression of $f(t)$ in \eqref{f(t)}, the right hand side of this equation can be cancelled provided that
\be
\gamma^\mu{}_{(\nu\rho)} = 2\Gamma^0{}_{\alpha(\nu}\frac{dx^\alpha}{dt}\delta^\mu_{\rho)}\,,
\ee
reducing the geodesic equation \eqref{GenGeodEq} to
\be
\frac{d^2x^\mu}{dt^2} + \hat\Gamma^\mu{}_{\nu\rho}\frac{dx^\nu}{dt}\frac{dx^\rho}{dt} = 0\,.\label{EquivGeodEq}
\ee
This shows that a reparametrization of \eqref{SimplGeodEq}, is equivalent to a transformation of the connection defined by
\be
\tilde\Gamma^\mu{}_{\nu\rho} = \Gamma^\mu{}_{\nu\rho} + 2\Gamma^0{}_{\alpha(\nu}\frac{dx^\alpha}{dt}\delta^\mu_{\rho)} + \gamma^\mu{}_{[\nu\rho]}\,.\label{ParConnTransf}
\ee
where $\gamma^\mu{}_{[\nu\rho]}$ carries 24 unconstrained degrees of freedom reminding that torsion does not affect the geodesic equation. Inspired by this symmetry, consider what is known as a projective transformation \cite{Eisenhart} that generalizes \eqref{ParConnTransf} and is defined as
\be
\tilde\Gamma^\mu{}_{\nu\rho} = \Gamma^\mu{}_{\nu\rho} + A_\nu\delta^\mu_\rho\,.\label{ProjTransf}
\ee
In terms of the tilded connection, the geodesic equation \eqref{GeodEq} reads
\be
\frac{d^2x^\mu}{dt^2} + \tilde\Gamma^\mu{}_{\nu\rho}\frac{dx^\nu}{dt}\frac{dx^\rho}{dt} = \left(f(t)+ A_\nu\frac{dx^\nu}{dt}\right)\frac{dx^\mu}{dt}\,,\label{GeodEqA}
\ee
so that the affine parameter associated to \eqref{GeodEqA} is defined as follows
\be
\frac{d\tau}{dt} = \tilde\xi \, e^{\int{f(t)dt}}\,,
\ee
where the integration constant $\tilde\xi$ is related to $\xi$ by a global rescaling 
\be
\tilde\xi = e^{\int A} \xi\,,
\ee
that depends on the one form $A = A_\mu dx^\mu$. Since a global rescaling of the integration constant $\xi$ does not affect the space of solutions of the geodesic equation \eqref{GeodEq}, the path of a test particles can be equivalently described by
\be
\frac{d^2x^\mu}{d\tau^2} + \tilde\Gamma^\mu{}_{\nu\rho}\frac{dx^\nu}{d\tau}\frac{dx^\rho}{d\tau} = 0\,,\label{ProjGeodEq}
\ee
This argument explains that projective invariance is a manifestation of the enhancement of symmetries associated to reparametrization invariance of the geodesic equation in the non-riemannian metric-affine geometry. As a consequance, it is possible to define an equivalence class of affine connections related by the projective transformation \eqref{ProjTransf}, that are ``projected'' into the same space of geodesic paths.

\vspace{0.5cm}

In order to understand how this symmetry acts on the action of a gravity theory, consider that the transformation rule of the Riemann tensor under a projective transformation \eqref{ProjTransf}, is 
\be
R^\alpha{}_{\beta\mu\nu}(\tilde \Gamma) = R^\alpha{}_{\beta\mu\nu}(\Gamma)+2\delta^\alpha_\beta\partial_{[\mu} A_{\nu]}\,.\label{RiemProjTransf}
\ee
where, following the metric-affine formalism, the Riemann tensor does not depend on the metric but just on the connection according to : 
\be
R^\alpha{}_{\beta\mu\nu} \equiv 2\delta^{\rho\sigma}_{\mu\nu}\left(\Gamma^\alpha{}_{\rho\lambda} + \delta^\alpha_\lambda\partial_\rho\right)\Gamma^{\lambda}{}_{\sigma\beta}\,,
\ee
The introduction of this compact way of writing the Riemann tensor in terms of the generalized Kronecker delta $\delta^{\mu\nu}_{\alpha\beta}=\displaystyle{\frac12\left(\delta^\mu_\alpha\delta^\nu_\beta - \delta^\mu_\beta\delta^\nu_\alpha\right)}$ is justified by the simplifications that occur in deriving \eqref{RiemProjTransf}.
As noticed in \cite{Bejarano:2019zco}, an immediate consequence of \eqref{RiemProjTransf} is that the Kretschmann scalar $K=R^\alpha{}_{\beta\mu\nu}R_\alpha{}^{\beta\mu\nu}$ is not invariant under projective transformations, putting forward the fundamental role of symmetries associated to the affine connection when dealing with curvature divergences in a metric affine framework.  
From \eqref{RiemProjTransf}, it follows immediately that the Ricci tensor transforms as
\be
R_{\beta\nu}(\tilde \Gamma) = R^\alpha{}_{\beta\mu\nu}(\tilde \Gamma)\delta^\mu_\alpha = R_{\beta\nu}(\Gamma)+2\partial_{[\alpha} A_{\beta]}\,.
\ee
When a Lagrangian depends on just the symmetric part of the Ricci tensor, the projective transformations become a true symmetry of the theory since
\be
R_{(\mu\nu)}(\tilde \Gamma) = R_{(\mu\nu)}(\Gamma)\,,
\ee
This symmetry enhancement implies that the class of gravity theories described by a Lagrangian that is an arbitrary function of the Ricci tensor, usually called Ricci-Based gravity (RBG), may have a special role in the metric-affine formalism. Indeed, this intuition is confirmed by the fact that for any RBG theory, it exists a redefinition of the metric that brings to a frame where the field equations associated to the connection are reduced to the metric compatibility condition up to a projective transformation. The original proof of this statement that has been provided in \cite{Afonso:2017bxr}, will be reviewed and simplified in the next section.

%%%%%%%%%%%%%%%%%%%%%%%%%%%%%%%%%%%%%%%%%%%%%%%%%%%%%%%%%%%%%%%%%
\section{Metric Compatibility of RBG Theories\label{MetricComp}}
%%%%%%%%%%%%%%%%%%%%%%%%%%%%%%%%%%%%%%%%%%%%%%%%%%%%%%%%%%%%%%%%%

The most general action of a RBG in the metric-affine formalism can be written as :
\be
S_{RB}=\int{d^4x \sqrt{|g|}\left[{\cal L}_G\left(g_{\mu\nu},R_{(\mu\nu)}\left(\Gamma\right)\right) + {\cal L}_m\left(g_{\mu\nu},\psi\right)\right]}\,,\label{RBG-Action}
\ee
where it is worth to stress that the Ricci tensor $R_{\mu\nu}\left(\Gamma\right)$ is a function of the connection alone and a matter Lagrangian ${\cal L}_m$ depending on bosonic fields $\psi$, has been included. The variation of this action with respect to the connection has been studied in detail in \cite{Afonso:2017bxr} resulting in the following field equations :
\be
\frac{1}{\sqrt{|g|}}\nabla_\mu\left(\sqrt{|g|} {\cal Z}^{\beta\lambda}\right) \delta^{\mu\nu}_{\alpha\lambda} = S^\nu{}_{\alpha\lambda}{\cal Z}^{\beta\lambda} + 2 S^\lambda{}_{\lambda\mu}{\cal Z}^{\beta\lambda}\delta^{\mu\nu}_{\alpha\lambda}\label{ConnFieldEq}
\ee
where it has been introduced the tensor
\be
{\cal Z}^{\mu\nu} \equiv \frac{\partial{{\cal L}_G}}{\partial R_{\mu\nu}}\,.\label{Z-Def}
\ee
Taking the trace of \eqref{ConnFieldEq} with respect to $\nu\alpha$, one finds the identity
\be
\frac{1}{2\sqrt{|g|}}\nabla_\nu\left(\sqrt{|g|} {\cal Z}^{\mu\nu}\right) = \frac23 S^\lambda{}_{\lambda\nu}{\cal Z}^{\mu\nu}\,,
\ee
that can be used to simplify \eqref{ConnFieldEq} into
\be
\frac{1}{2\sqrt{|g|}}\nabla_\lambda\left(\sqrt{|g|} {\cal Z}^{\mu\nu}\right)  = -\frac13 S^\alpha{}_{\alpha\beta}{\cal Z}^{\mu\beta}\delta^\nu_\lambda + S^\nu{}_{\lambda\alpha}{\cal Z}^{\mu\alpha} + S^\alpha{}_{\alpha\alpha}{\cal Z}^{\mu\nu}\,.\label{ConnFieldEqTraced}
\ee
Following the standard procedure to remove the $\sqrt{|g|}$ factor inside the covariant derivative and assuming that the inverse of \eqref{Z-Def} ${\cal Z}_{\mu\nu}$ exists, the last equation can be multiplied by ${\cal Z}_{\mu\nu}$ to obtain the following expression of the trace of the connection :
\be
\Gamma^\alpha{}_{\mu\alpha} = \partial_\mu\ln{\left(\frac{|g|}{\sqrt{|Z|}}\right)} - \frac83 S^\alpha{}_{\alpha\mu}\,.
\ee
This identity can be used to simplify \eqref{ConnFieldEqTraced} into
\be
\left(\nabla_\lambda - \partial_\alpha\sqrt{\left|\frac{g}{\cal Z}\right|}\right){\cal Z}^{\mu\nu}  = -\frac43 S^\alpha{}_{\alpha\beta}{\cal Z}^{\mu\gamma}\delta^\beta_{(\lambda}\delta^\nu_{\gamma)} + 2 S^\nu{}_{\lambda\alpha}{\cal Z}^{\mu\alpha}\,.\label{ConnFieldEqSimpl}
\ee
Introducing the following tensor
\be
q^{\mu\nu} \equiv  \xi\sqrt{\left|\frac{\cal Z}{g}\right|} {\cal Z}^{\mu\nu}\,,\label{qDef}
\ee
where $\xi$ is an arbitrary constant factor, equation \eqref{ConnFieldEqSimpl} can be further simplified into
\be
\nabla_\lambda {q}^{\mu\nu}  = -\frac43 S^\alpha{}_{\alpha\beta}q^{\mu\gamma}\delta^\beta_{(\lambda}\delta^\nu_{\gamma)} + 2 S^\nu{}_{\lambda\alpha}q^{\mu\alpha}\,.\label{ConnFieldEqMin}
\ee
This form of the field equations of the connection is the closest to the standard metric compatibility condition that can be obtained using algebraic manipulations and a redefinition of the metric. At this point, projective transformation comes into the game. In fact, shifting the connection according to
\be
\Gamma^\lambda_{\mu\nu} = \tilde\Gamma^\lambda_{\mu\nu} - \frac23 S^\alpha_{\mu\alpha} \delta^\lambda_\nu\,,\label{SpecProjTransf}
\ee
that is a specific form of the projective transformation \eqref{ProjTransf} with $A_\mu =\displaystyle{ \frac23 }S^\alpha{}_{\mu\alpha}$, the field equations \eqref{ConnFieldEqMin} assume their final form
\be
\tilde\nabla_\lambda {q}^{\mu\nu}  = 2 \tilde S^\nu{}_{\lambda\alpha}q^{\mu\alpha}\,.\label{FinConnEq}
\ee

Since any RBG depend on a symmetric Ricci tensor by definition, the tensor $q_{\mu\nu}$, defined through \eqref{Z-Def} and \eqref{qDef} is symmetric. Reminding this property and summing permutations of \eqref{FinConnEq} with proper signs and lowered indices, leads to the following identity :
\be
\partial_\alpha q_{\mu\nu} + \partial_\nu q_{\alpha\mu} - \partial_\mu q_{\nu\alpha} = 2 \tilde\Gamma^\lambda{}_{\alpha\nu}q_{\lambda\mu}\,,
\ee
that, symmetrized and antisymmetrized in $\alpha\nu$, says that the symmetric part of the connection coincides with the Christoffel symbol associated to the auxiliary metric 
\be
\tilde\Gamma^\lambda{}_{(\alpha\nu)} = \frac12 q^{\lambda\mu}\left(\partial_\alpha q_{\mu\nu} + \partial_\nu q_{\alpha\mu} - \partial_\mu q_{\nu\alpha}\right)\,,\label{MetrComp}
\ee
while torsion vanishes in the tilded frame, namely $\tilde S^\lambda{}_{\alpha\nu}=0$. In light of \eqref{MetrComp}, the tensor $q_{\mu\nu}$ can be regarded as the proper auxiliary metric that describe a Riemannian frame. Its definition can be further elaborated by taking the determinant of \eqref{qDef} so that the following relation can be derived
\be
\sqrt{|q|} = \frac{1}{\xi^2}\frac{\left|g\right|}{\sqrt{\left|{\cal Z}\right|}}\,,
\ee
that simplifies \eqref{qDef} as follows :
\be
\sqrt{|q|}q^{\mu\nu} = \frac{1}{\xi}\sqrt{|g|}Z^{\mu\nu}\,.\label{qDefSimpl}
\ee
Notice that the constant factor $\xi$ introduced in \eqref{qDef} parametrizes a freedom in defining $q_{\mu\nu}$ that is not constrained by the connection field equations. This observation will be relevant to simplify the field equations associated to the variation of the metric in the next section.

%%%%%%%%%%%%%%%%%%%%%%%%%%%%%%%%%%%%%%%%%%%%%%%%%%%%%%%%%%%%%%%%%
\section{Correspondence Between RBG Theories and GR\label{Corr-RBG-GR}}
%%%%%%%%%%%%%%%%%%%%%%%%%%%%%%%%%%%%%%%%%%%%%%%%%%%%%%%%%%%%%%%%%

The possibility of defining a frame where the projectively transformed connection is the Christoffel symbol of an auxiliary metric, paves the way for an on-shell correspondence between RBG and GR. In fact, the field equations associated to the connection of any RBG and GR are the same up to a redefinition of the metric \eqref{qDefSimpl} and a projective transformation \eqref{SpecProjTransf}. A natural question is whether this equivalence can be extended to the metric field equation by modifying the matter sector in some way. In order to answer this question, consider that the variation of the action \eqref{RBG-Action} with respect to the metric gives the following field equations
\be
g^{\mu\rho}\frac{\delta {\cal L}_G}{\delta g^{\rho\nu}} - \frac12 {\cal L}_G \delta^\mu_\nu = \frac12 T^\mu{}_\nu\,,\label{MetrFieldEqs}
\ee
where the energy-momentum tensor is defined in the usual way
\be
T_{\mu\nu} = -\frac{2}{\sqrt{|g|}}\frac{\delta \left(\sqrt{|g|}{\cal L}_m\right)}{\delta g^{\mu\nu}}\,.\label{EnMomTens}
\ee
Diffeomorphic invariance of the action implies that the Lagrangian depends on the Ricci tensor only through the combination $R^\mu{}_\nu = g^{\mu\rho}R_{\rho\nu}$. Applying the chain rule, this dependence implies the relations
\be
\frac{\delta {\cal L}_G\left(R^\mu{}_\nu\right)}{\delta g^{\alpha\beta}} = \frac{\delta {\cal L}_G\left(R^\mu{}_\nu\right)}{\delta R^\alpha{}_\rho}R_{\beta\rho}\,,
\ee
\be
\frac{\delta {\cal L}_G\left(R^\mu{}_\nu\right)}{\delta R_{\alpha\beta}} = \frac{\delta {\cal L}_G\left(R^\mu{}_\nu\right)}{\delta R^\lambda{}_\beta}g_{\lambda\alpha}\,.
\ee
It follows immediately the following identity
\be
\frac{\delta {\cal L}_G\left(R^\mu{}_\nu\right)}{\delta g^{\rho\sigma}} = \frac{\delta {\cal L}_G\left(R^\mu{}_\nu\right)}{\delta R_{\alpha\beta}}g_{\alpha\rho}R_{\beta\sigma}\,,
\ee 
that can be used to replace the variation with respect to the metric with the one with respect to the Ricci tensor in \eqref{MetrFieldEqs} to obtain
\be
\frac{\delta{\cal L}_G}{\delta R_{\mu\rho}} R_{\rho\nu}(\Gamma) = \frac12 T^\mu{}_\nu + \frac12 {\cal L}_G\delta^\mu_\nu\,.\label{PreEinstein}
\ee
Turning the attention to the GR side, the Einstein-Hilbert action coupled to matter that will be considered is
\be
S_{EH} = \int \sqrt{-q}\left[\frac{ q^{\mu\nu}R_{\mu\nu}\left(\tilde\Gamma\right)}{2\kappa^2} + \tilde {\cal L}_m\left(\psi,q\right)\right]\label{EinsteinHilbert}
\ee
where $\kappa^2 =8\pi G/c^4 $ is the usual coupling constant of GR and the matter sector is described by the Lagrangian $\tilde{\cal L}_m$. Notice that this action is written in terms of the auxiliary metric $q_{\mu\nu}$ defined in \eqref{qDefSimpl} to guarantee the correspondence between the connection field equations in the GR and RBG frames.
The Einstein field equations obtained from the variation of this action with respect to the metric, can be equivalently written as follows
\be
q^{\mu\rho} R_{\rho\nu}(\tilde\Gamma )=\kappa^2\left(\tilde T^\mu{}_\nu-\frac12 \tilde T \delta^\mu_\nu\right)\,,\label{AnalogueGR}
\ee
where the energy-momentum tensor has the same functional form of \eqref{EnMomTens} but in terms of fields in the GR frame, namely 
\be
\tilde T_{\mu\nu} = -\frac{1}{\sqrt{|q|}}\frac{\delta \left(\sqrt{|q|}\tilde{\cal L}_m\right)}{\delta q_{\mu\nu}}\,,\label{EnMomTens}
\ee
with $\tilde T$ being its trace. Taking into account the definition of the auxiliary metric \eqref{qDefSimpl}, the field equations \eqref{PreEinstein} and \eqref{AnalogueGR} are equivalent if the following relations hold : 
\be
\sqrt{-q} q^{\mu\nu} = 2\kappa^2\sqrt{-g} \frac{\delta{\cal L}_G}{\delta R_{\mu\nu}}\,,\label{AuxMetrEq}
\ee
\be
\sqrt{\left|q\right|}\left(\tilde T^\mu{}_\nu-\frac12 \tilde T \delta^\mu_\nu\right) = \sqrt{\left|g\right|}\left( T^\mu{}_\nu + {\cal L}_G\delta^\mu_\nu\right)\,.\label{PreCond}
\ee
Indeed, equation \eqref{AuxMetrEq} fixes the unspecified constant in \eqref{qDef} according to $\xi = 1/(2\kappa^2)$ and reduces \eqref{PreEinstein} to
\be
q^{\mu\rho} R_{\rho\nu}(\tilde\Gamma) = \kappa^2\sqrt{\left|\frac{g}{q}\right|}\left( T^\mu{}_\nu + {\cal L}_G\delta^\mu_\nu\right) \,.\label{AlmostEinstein}
\ee
where the projective transformation \eqref{SpecProjTransf}, that leaves invariant the Ricci tensor, has been applied. The on-shell correspondence between a RBG and GR is complete if the right hand sides of \eqref{AnalogueGR} and \eqref{AlmostEinstein} coincide, giving rise to the condition \eqref{PreCond}. Taking the trace of the equation \eqref{PreCond}, one finds
\be
\sqrt{\left|g\right|}T  + \sqrt{\left|q\right|}\tilde T  = - 4\sqrt{\left|g\right|} {\cal L}_G \,,\label{TraceCond}
\ee
so that the energy-momentum tensor in the GR frame can be written in terms of matter fields defined in the RBG theory as follows :
\be
\sqrt{\left|q\right|}\,\tilde T^\mu{}_\nu = \sqrt{\left|g\right|}\left(T^\mu{}_\nu - \frac12 T\delta^\mu_\nu - {\cal L}_G\delta^\mu_\nu\right)\,.\label{Tilde-T-Corr}
\ee
where it is important to underline that ${\cal L}_G$ is calculated on a specific solution of the RBG.  This equation makes manifest that the net effect of the mapping is to shift nonlinearities on the gravity sector of a RBG into the matter sector coupled to GR. Combining \eqref{PreCond} with \eqref{TraceCond}, it is possible to find the energy-momentum tensor of the RBG theory in terms of matter fields associated to the GR frame according to :
\be
\sqrt{\left|g\right|}\,T^\mu{}_\nu = \sqrt{\left|q\right|}\left(\tilde T^\mu{}_\nu - \frac12 \tilde T\delta^\mu_\nu\right) - \sqrt{\left|g\right|}\,{\cal L}_G\delta^\mu_\nu\,.\label{T-Corr}
\ee
As in the previous case, ${\cal L}_G$ is intended as a function of matter fields since it is calculated on a solution of the GR theory. Resuming, the following statement holds true : using the metric-affine formalism, any RBG admits a framewith associated metric $q_{\mu\nu}$ defined by \eqref{AuxMetrEq}, where the space of solutions solve the Einstein field equations provided that the matter sector interactions are redefined according to the system of differencial equations \eqref{PreCond}. 

%For the sake of pratical applications, eq

%%%%%%%%%%%%%%%%%%%%%%%%%%%%%%%%%%%%%%%%%%%%%%%%%%%%%%%%%%%%%%%%%
\section{Generating RBG Solutions\label{GenerSolut}}
%%%%%%%%%%%%%%%%%%%%%%%%%%%%%%%%%%%%%%%%%%%%%%%%%%%%%%%%%%%%%%%%%

In this section will be outlined a general approach and its limitations to find a solution $g_{\mu\nu}$ of the field equations associated to a given RBG described by ${\cal L}_G$, using a known solution $q_{\mu\nu}$ of GR. In this setting, the matter sector in the GR frame ${\cal L}_m$, is not known a priori since it will be provided by the machinery developed in the next section, eventually depending on the choice of ${\cal L}_G$ and $\tilde{\cal L}_m$. 

The starting point is the definition \eqref{AuxMetrEq} from which it is straightforward to find the auxiliary metric in terms of the RBG metric, namely
\be
q^{\mu\nu} = \frac{1}{2\kappa^2\sqrt{\det{\left(\frac{\delta{\cal L}_G}{\delta R^\rho{}_{\sigma}}\right)}}}\frac{\delta{\cal L}_G}{\delta R_{\mu\nu}}\,,\label{q-g-map}
\ee
that in general is a nonlinear function of $g_{\mu\nu}$. Since ${\cal L}_G$ is given, this formula provide an expression of the inverse auxiliary metric associated to a given RBG metric. Since the solution of a RBG theory is usually more difficult to find due to the possible presence of nonlinearities in the gravity sector, this result is not as appealing as the inverse problem, namely when the auxiliary metric is known and the RBG solution has to be found. 

Assuming that the auxiliary metric is related to the original metric in the RBG frame by a linear transformation, crucial simplifications allow to deal with the inverse problem. In fact, introducing the deformation matrix ($\Omega$-matrix for short) as follows
\be
q^{\mu\nu} = g^{\mu\rho}\left(\Omega^{-1}\right)_\rho{}^\nu\,,\label{Om-Def}
\ee
the inverse problem is reduced to finding the $\Omega$-matrix in terms of the GR fields so that the RBG solution will be generated by the equation
\be
g_{\mu\nu} = \left(\Omega^{-1}\right)_\mu{}^\rho q_{\rho\nu}\,.
\ee
From the definition \eqref{Om-Def} and \eqref{q-g-map}, it steadily follows that the deformation matrix in terms of RBG fields is given by
\be
\left(\Omega^{-1}\right)^\nu{}_\mu = \frac{1}{2\kappa^2\sqrt{-\det{\left(\frac{\delta{\cal L}_G}{\delta R^\alpha{}_{\beta}}\right)}}}\frac{\delta{\cal L}_G}{\delta R^\mu{}_{\nu}}\,.\label{Om-RBG}
\ee
that is another way of writing
\be
\frac{\delta{\cal L}_G}{\delta R^\mu{}_{\nu}} = \frac{1}{2\kappa^2}\sqrt\Omega\left(\Omega^{-1}\right)^\nu{}_\mu \,,\label{R-Om}
\ee
where the determinant of the deformation matrix $\Omega$ has been introduced. This confirms that it is always possible to introduce a deformation matrix if the right hand side of \eqref{Om-RBG} is well defined. However, in order to generate RBG solutions, one needs to find a way to write \eqref{Om-RBG} in terms of quantities in the GR frame. The relation between the $\Omega$-matrix and the GR fields is encoded in the metric field equation \eqref{PreEinstein} that can be cast as follows 
\be
\left(\Omega^{-1}\right)^\mu{}_\rho R^\rho{}_\nu(\tilde\Gamma )=\kappa^2\left(\tilde T^\mu{}_\nu-\frac12 \tilde T \delta^\mu_\nu\right)\,,\label{Omega-Tilde-T}
\ee
Unfortunately, this equation is not useful unless $R^\rho{}_\nu(\tilde\Gamma )=q^{\rho\sigma}R_{\sigma\nu}\left(\Gamma\right)$ is expressed as a function of GR fields. This is the main obstruction to generalize the inverse mapping between metrics in different frames to any theory, leaving the only possibility of a case by case study. In some cases, a possible way out comes from solving \eqref{R-Om} to find the Ricci tensor $R^\rho{}_\nu$ in terms of the deformation matrix. Once substituted, equation \eqref{Omega-Tilde-T} boil down to an algebraic  equation for the $\Omega$-matrix in terms of GR fields. An important instance where this strategy can be applied, is represented by the Eddington-inspired-Born-Infeld gravity that will be worked out in section \ref{EiBIsection}.

%%%%%%%%%%%%%%%%%%%%%%%%%%%%%%%%%%%%%%%%%%%%%%%%%%%%%%%%%%%%%%%%%
\section{Correspondence Between Matter Sectors\label{CorrMattSect}}
%%%%%%%%%%%%%%%%%%%%%%%%%%%%%%%%%%%%%%%%%%%%%%%%%%%%%%%%%%%%%%%%%

Having in mind the problem of finding a RBG solution that corresponds to a given GR metric, the matter sector ${\cal L}_m$ is not known a priori but, as mentioned at the beginning of the previous section, it can be unveiled once the RBG theory ${\cal L}_G$ is chosen. The relation between the matter sectors is encoded in the condition \eqref{PreCond} that explicitly reads
\bea
&&\displaystyle{\sqrt{|g|}\left[\delta^\mu_\nu\left(\tilde{\mathcal{L}}_m(q,\psi) - {q}^{\rho\sigma}\frac{\delta\tilde{\cal L}_m(q,\psi)}{\delta {q}^{\rho\sigma}}\right)+2 {q}^{\mu\rho}\frac{\delta\tilde{\cal L}_m(q,\psi)}{\delta {q}^{\rho\nu}}\right]=}\\
&&\qquad\qquad\qquad\qquad\qquad\qquad\qquad\qquad\displaystyle{=\sqrt{|q|} \left[2g^{\mu\rho}\frac{\delta{\cal L}_m(g,\psi)}{\delta g^{\rho\nu}} - \delta^\mu_\nu\left(\mathcal{L}_G + \mathcal{L}_m(g,\psi)\right)\right]}\,,\label{mapping}
\eea
since the energy-momentum tensor in terms of the Lagrangian is written as
\be
{T^\mu}_\nu=g^{\mu\alpha}T_{\alpha\nu}={\cal L}_m(g_{\mu\nu},\psi)\delta^\mu{}_\nu-2g^{\mu\rho}\frac{\delta{\cal L}_m(g_{\mu\nu},\psi)}{\delta g^{\rho\nu}}\,.\label{T0}
\ee
Once again, the trace of \eqref{mapping} provide the relation
\be
\sqrt{|q|}\left(\mathcal{L}_m(g,\psi) - \frac12 g^{\mu\nu}\frac{\delta{\cal L}_m(g,\psi)}{\delta g^{\mu\nu}} +\mathcal{L}_G\right) = \sqrt{|g|}\left(\frac12q^{\mu\nu}\frac{\delta\tilde{\cal L}_m(q,\psi)}{\delta q^{\mu\nu}} - \tilde{\cal L}_m(q,\psi)\right)\,,\label{TrCond}
\ee
that simplifies \eqref{mapping} into
\be
\sqrt{|q|}\left(\frac12{q}^{\rho\sigma}\frac{\delta\tilde{\cal L}_m(q,\psi)}{\delta {q}^{\rho\sigma}} \delta^\mu_\nu -2 {q}^{\mu\rho}\frac{\delta\tilde{\cal L}_m(q,\psi)}{\delta {q}^{\rho\nu}}\right) = \sqrt{|g|}\left(\frac12g^{\rho\sigma}\frac{\delta{\cal L}_m(g,\psi)}{\delta g^{\rho\sigma}}\delta^\mu_\nu - g^{\mu\rho}\frac{\delta{\cal L}_m(g,\psi)}{\delta g^{\rho\nu}}\right)\,.\label{TrLessCond}
\ee
This means that the traceless tensor
\be
I^\mu{}_\nu = \sqrt{|g|}\left(\frac12g^{\rho\sigma}\frac{\delta{\cal L}_m(g,\psi)}{\delta g^{\rho\sigma}} \delta^\mu_\nu- g^{\mu\rho}\frac{\delta{\cal L}_m(g,\psi)}{\delta g^{\rho\nu}}\right)\,.
\ee
is invariant under the mapping between the RBG and GR. Finally, the most general equations that define the correspondence between the matter Lagrangians in the two frames are \eqref{TrCond} and \eqref{TrLessCond}. A simplified version of these equations can be obtained by imposing the condition
\be
\sqrt{|g|}g^{\mu\rho}\frac{\delta{\cal L}_m(g,\psi)}{\delta g^{\rho\nu}} = \sqrt{|q|}{q}^{\mu\rho}\frac{\delta\tilde{\cal L}_m(q,\psi)}{\delta {q}^{\rho\nu}}\,,\label{Restr}
\ee
that is a particular case of \eqref{TrLessCond} used in \cite{Afonso:2018bpv,Afonso:2018hyj,Afonso:2018mxn,Delhom:2019zrb}. The advantage of restricting the space of solutions is that the condition \eqref{TrCond} simplifies to provide the following parametrizations of the matter Lagrangian of the RBG and GR frames  
\be
\sqrt{|g|}\mathcal{L}_m(g,\psi) = \sqrt{|q|}\left(q^{\mu\nu}\frac{\delta\tilde{\cal L}_m(q,\psi)}{\delta q^{\mu\nu}} - \tilde{\cal L}_m(q,\psi)\right)-\sqrt{|g|}\mathcal{L}_G\,,\label{Paramtr}
\ee
\be
\sqrt{|q|}\tilde{\cal L}_m(q,\psi) = \sqrt{|g|}\left(\mathcal{L}_m(g,\psi) - \frac12 g^{\mu\nu}\frac{\delta{\cal L}_m(g,\psi)}{\delta g^{\mu\nu}} +\mathcal{L}_G\right)\,.\label{GRParamtr}
\ee
Even though these equations represent a progress in providing a general recipe to find the matter counterpart associated to a different frame, the right hand sides are still written in terms of the auxiliary metric and the gravity Lagrangian of the RBG in the first case and in terms of the GR metric in the second. Concerning the inverse problem, the strategy to follow to find the functional form of $\mathcal{L}_m(g,\psi)$ from \eqref{Paramtr} depends on the type of matter content of the theory. In particular, it is needed a mechanism that allows to express the gravity Lagrangian in the right hand side of \eqref{Paramtr} as a function of matter fields. Same argument applies to find $\tilde{\cal L}_m(q,\psi)$ from \eqref{GRParamtr}. The case of an arbitrary number of scalar fields has been studied in a series of papers \cite{Afonso:2018bpv,Afonso:2018hyj} while the case of an electric field can be found in \cite{Afonso:2018bpv}, later generalized to the electromagnetic case in \cite{Delhom:2019zrb}.

%%%%%%%%%%%%%%%%%%%%%%%%%%%%%%%%%%%%%%%%%%%%%%%%%%%%%%%%%%%%%%%%%
\section{Correspondence between the EiBI theory and GR\label{EiBIsection}}
%%%%%%%%%%%%%%%%%%%%%%%%%%%%%%%%%%%%%%%%%%%%%%%%%%%%%%%%%%%%%%%%%

The general approach to find the correspondence between GR and RBG described in the previous sections, is particularly effective in the case of the EiBI theory \cite{Banados:2010ix} where the gravity Lagrangian ${\cal L}_{G}$ is defined as
\be
{\cal L}_{EiBI} = \frac{1}{\epsilon\kappa^2}\left[\sqrt{\det{\left(\delta^\mu_\nu +\epsilon R^\mu{}_\nu\right)}} - \lambda\right]\,.
\ee
Here, the lenght-squared dimensional parameter $\epsilon$ controls nonlinearity while the parameter $\lambda$ produces a cosmological constant in the $\epsilon$-expansion of the Lagrangian (see \cite{BeltranJimenez:2017doy} for a review). 
Since
\be
\frac{\delta{\cal L}_{EiBI}}{\delta R^{\mu}{}_{\nu}} = \frac{1}{2\kappa^2}\sqrt{\det{\left(\delta^\rho{}_\sigma+\epsilon R^\rho{}_\sigma\right)}}\left[\left(\hat I+\epsilon \hat R\right)^{-1}\right]_\mu{}^{\nu}\,,
\ee
the equation \eqref{R-Om} leads to the well known deformation matrix associated to the EiBI gravity theory
\be
\Omega_\nu{}^\mu = \delta^\mu_\nu+\epsilon R^{\mu}{}_\nu\,.\label{Om-R}
\ee
This relation reveals a simple dependence of the Ricci tensor with respect to the $\Omega$-matrix, namely
\be
R^{\mu}{}_\nu\left(\Omega\right) = \frac{1}{\epsilon}\left(\Omega^\mu{}_\nu - \delta^\mu_\nu\right)\,.\label{R-Omega}
\ee
Substituting \eqref{R-Omega} in the field equations \eqref{Omega-Tilde-T}, one finds the deformation matrix in terms of the GR fields : 
\be
\left(\Omega^{-1}\right)^\mu{}_\nu = \delta^\mu_\nu - \epsilon\kappa^2\left(\tilde T^\mu{}_\nu-\frac12 \tilde T \delta^\mu_\nu\right)\,,\label{InvOmFin}
\ee
finally leading to the RBG metric generated by the GR metric :
\be
g_{\mu\nu} = q_{\mu\nu} - \epsilon\kappa^2\left(\tilde T_{\mu\nu}-\frac12 \tilde T q_{\mu\nu}\right)\,,\label{RBGMetr}
\ee
Concerning the correspondence between the matter sectors, a peculiar feature of the EiBI theory is that the Lagrangian can be written in terms of the deformation matrix as follows
\be
{\cal L}_{EiBI}\left(\Omega\right) = \frac{\sqrt{\Omega}-\lambda}{\epsilon \kappa^2}\,,\label{EiBI-Om}
\ee
so that, using \eqref{R-Omega}, the metric field equations \eqref{PreEinstein} of the RBG frame become 
\be
 \sqrt{\Omega}\left(\Omega^{-1}\right)_\nu{}^\mu = \lambda\delta^\mu_\nu - \epsilon\kappa^2 T^\mu{}_\nu \,,\label{Om-T}
\ee
thus providing the deformation matrix
\be
\left(\Omega^{-1}\right)_\nu{}^\mu =\displaystyle{\frac{\lambda\delta^\mu_\nu - \epsilon\kappa^2 T^\mu{}_\nu}{ \sqrt{\det{\left(\lambda\delta^\rho_\sigma - \epsilon\kappa^2 T^\rho{}_\sigma\right)}}}} \,,\label{OmRBG}
\ee
and the GR metric
\be
q^{\mu\nu} = g^{\mu\rho}\left(\Omega^{-1}\right)_\rho{}^\mu =\displaystyle{\frac{\lambda g^{\mu\nu} - \epsilon\kappa^2 T^{\mu\nu}}{ \sqrt{\det{\left(\lambda\delta^\rho_\sigma - \epsilon\kappa^2 T^\rho{}_\sigma\right)}}}} \,,\label{GRMetr}
\ee
as functions of RBG fields.
The last ingredient that is needed to write the right hand side of \eqref{Paramtr} in the RBG frame is the gravity sector ${\cal L}_G$ that, in light of \eqref{EiBI-Om} and  \eqref{OmRBG}, can be written in terms of matter fields as follows
\be
{\cal L}_{EiBI} = \frac{1}{\epsilon\kappa^2}\left[\sqrt{\det{\left(\lambda\delta^\rho_\sigma - \epsilon\kappa^2 T^\rho{}_\sigma\right)}}-\lambda\right]\,.\label{RBGMatterEiBI}
\ee
Alternatively, the EiBI Lagrangian can be written in terms of GR fields according to
\be
{\cal L}_{EiBI} = \frac{1-\lambda \sqrt{\det{\left[\delta^\rho_\sigma - \epsilon\kappa^2 \left(\tilde T^\rho{}_\sigma-\frac12\tilde T \delta^\rho_\sigma\right)\right]}}}{\epsilon\kappa^2\sqrt{\det{\left[\delta^\rho_\sigma - \epsilon\kappa^2 \left(\tilde T^\rho{}_\sigma-\frac12\tilde T \delta^\rho_\sigma\right)\right]}}}\,,\label{GRMatterEiBI}
\ee
as a consequence of \eqref{EiBI-Om} and \eqref{InvOmFin}. Upon substituting the gravity Lagrangian \eqref{RBGMatterEiBI} and \eqref{GRMatterEiBI} in the parametrizations \eqref{Paramtr} and \eqref{GRParamtr}, one finds the following parametrizations of the matter sectors
\be
\mathcal{L}_m(g,\psi) = \left.\frac{1}{\epsilon\kappa^2}\left\{\lambda-\frac{1-\epsilon\kappa^2\left(\tilde{\cal L}_m-\frac12\tilde T\right)}{\sqrt{\det{\left[\delta^\rho_\sigma - \epsilon\kappa^2 \left(\tilde T^\rho{}_\sigma-\frac12\tilde T \delta^\rho_\sigma\right)\right]}}}\right\}\right|_{q=q(g)}\label{RBG-Lag}
\ee
\be
\tilde{\cal L}_m(q,\psi) = \left.\frac{1}{\epsilon\kappa^2}\left[\frac{\lambda + \epsilon\kappa^2 \left({\cal L}_m - \frac12 T \right)}{\sqrt{\det{\left(\lambda\delta^\rho_\sigma - \epsilon\kappa^2 T^\rho{}_\sigma\right)}}} - 1\right]\right|_{g=g(q)}\,.\label{GR-Lag}
\ee
where $q(g)$ and $g(q)$ can be obtained inverting the relations \eqref{RBGMetr} and \eqref{GRMetr}, respectively.
%More explicitly
%\be
%\mathcal{L}_m(g,\psi) = \frac{\lambda}{\epsilon\kappa^2} - \left.\frac{1-\epsilon\kappa^2\left(q^{\mu\nu}\frac{\delta\tilde{\cal L}_m(q,\psi)}{\delta q^{\mu\nu}} - \tilde{\cal L}_m(q,\psi)\right)}{\epsilon\kappa^2\sqrt{\det{\left\{\left[1 - \epsilon\kappa^2 \left(q^{\mu\nu}\frac{\delta\tilde{\cal L}_m(q,\psi)}{\delta q^{\mu\nu}} - \tilde{\cal L}_m(q,\psi)\right)\right]\delta^\rho_\sigma + 2 \epsilon\kappa^2q^{\rho\mu}\frac{\delta\tilde{\cal L}_m(q,\psi)}{\delta q^{\mu\sigma}}\right\}}}}\right|_{q=q(g)}
%\ee
%\be
%\tilde{\cal L}_m(q,\psi) = \left.\frac{\lambda + \epsilon\kappa^2 \left( g^{\mu\nu}\frac{\delta{\cal L}_m(g,\psi)}{\delta g^{\mu\nu}} -\mathcal{L}_m(g,\psi) \right)}{\epsilon\kappa^2\sqrt{\det{\left[\lambda\delta^\rho_\sigma + \epsilon\kappa^2 \left( g^{\rho\mu}\frac{\delta{\cal L}_m(g,\psi)}{\delta g^{\mu\sigma}} -\mathcal{L}_m(g,\psi)\delta^\rho_\sigma\right)\right]}}}\right|_{g=g(q)} - \frac{1}{\epsilon\kappa^2}\,.
%\ee
It is worth to stress that \eqref{RBG-Lag} and \eqref{GR-Lag} provide a generalization of the correspondence for scalars and electromagnetic fields as given in \cite{Afonso:2018hyj,Delhom:2019zrb} to any bosonic matter content. However, solving the equations \eqref{RBGMetr} and \eqref{GRMetr} with respect to $q_{\mu\nu}$ and $g_{\mu\nu}$ respectively, could not be immediate and require a case by case study. In the case of gauge fields, a possible approach is based on the fact that the matter Lagrangian depends on the metric only through invariants of the gauge group. Therefore, the replacement of the metric in the right hand side of \eqref{RBG-Lag} and \eqref{GR-Lag} can be understood as a map of the gauge invariants between different frames. This intuition played a fundamental role in establishing the correspondence in the case of electromagnetic fields \cite{Delhom:2019zrb} as will be reviewed in the next section.  

%%%%%%%%%%%%%%%%%%%%%%%%%%%%%%%%%%%%%%%%%%%%%%%%%%%%%%%%%%%%%%%%%
\section{Generating Solutions for EiBI coupled to Electromagnetic Fields\label{GenSol}}
%%%%%%%%%%%%%%%%%%%%%%%%%%%%%%%%%%%%%%%%%%%%%%%%%%%%%%%%%%%%%%%%%

When the correspondence is applied to matter sectors that describe a $U(1)$ gauge fields theory, the fundamental intuition that allows to write explicitly the right hand side of \eqref{RBG-Lag} and \eqref{GR-Lag} is that the theory  is based on a Lagrangian that is a function of just two electromagnetic invariants defined as  $K=\frac12 F_{\mu\nu}F^{\nu\mu}$ and $G=\frac14 F_{\mu\nu}{}^\star F^{\mu\nu}$. This is a consequence of the reduction property of arbitrary products of the field strenghts that can be stated as follows  
\be
F^{\mu}{}_{\lambda_1}F^{\lambda_1}{}_{\lambda_2}\cdots F^{\lambda_{p-2}}{}_{\lambda_{p-1}} F^{\lambda_{p-1}}{}_{\lambda_p}=
	F^\mu{}_{\lambda_1}F^{\lambda_1}{}_{\lambda_2}\cdots F^{\lambda_{p-5}}{}_{\lambda_p}	
	G^2 +
	F^\mu{}_{\lambda_1}F^{\lambda_1}{}_{\lambda_2}\cdots F^{\lambda_{p-3}}{}_{\lambda_{p-2}}K\,,
\ee
that , in turn, is based on the multiplication rules of the field strenghts
\begin{equation}
F^\mu{}_\lambda F^{\lambda}{}_\nu={}^\star F^\mu{}_\lambda {}^\star F^\lambda{}_\nu + \delta^\mu_\nu K
\,,\qquad {}^\star F^\mu{}_\lambda F^\lambda{}_\nu = F^\mu{}_\lambda {}^\star F^\lambda{}_\nu = - \delta^\mu_\nu G\,.\label{BasicRel}
\end{equation}
It steadily follows that the derivatives of the matter Lagrangian with respect to the metric can be written in terms of derivatives with respect to the electromagnetic invariants according to
\be
\frac{\delta{\cal L}_m}{\delta g^{\mu\nu}} = \frac{\partial{\cal L}_m}{\partial K}K_{\mu\nu} + \frac12\frac{\partial{\cal L}_m}{\partial G}\,G\,g_{\mu\nu}
\ee
where the following tensor quantity has been introduced
\be
K_{\mu\nu}=\frac{\partial K}{\partial g^{\mu\nu}} = F_{\mu\rho}F^\rho{}_\nu\,.
\ee
The K-tensor just defined obey the multiplication rule
\be
K^\mu{}_\rho K^\rho{}_\nu = G^2\delta^\mu_\nu + K\,K^\mu{}_\nu\,,\label{KK}
\ee
that implies a closure under multiplication of the identity and $K_{\mu\nu}$. Following these observations, the correspondence between the metrics in presence of arbitrary matter fields given by \eqref{RBGMetr} and \eqref{GRMetr}, can be specified to the electromagnetic case as follows
\be
g_{\mu\nu} = \left[1+\epsilon\kappa^2\left(\tilde{\cal L}_m - 2\tilde K\frac{\partial \tilde{\cal L}_m}{\partial\tilde K} - \tilde G\frac{\partial \tilde{\cal L}_m}{\partial\tilde G}\right)\right]q_{\mu\nu} + 2 \epsilon\kappa^2 \frac{\partial \tilde{\cal L}_m}{\partial\tilde K}\tilde K_{\mu\nu} \,,\label{EiBI-g-q}
\ee
\be
q_{\mu\nu} = \left[\lambda-\epsilon\kappa^2\left({\cal L}_m - 2K\frac{\partial {\cal L}_m}{\partial K} - G\frac{\partial {\cal L}_m}{\partial G}\right)\right]g_{\mu\nu} -  2\epsilon\kappa^2 \frac{\partial {\cal L}_m}{\partial K} K_{\mu\nu}\label{EiBI-q-g}
%q^{\mu\nu} = \frac{\left[\lambda+\epsilon\kappa^2\left(G\frac{\partial {\cal L}_m}{\partial G}-{\cal L}_m\right)\right]\delta^\mu_\rho + 2\epsilon\kappa^2 \frac{\partial {\cal L}_m}{\partial K} K^\mu{}_\rho}{\sqrt{\det\left\{\left[\lambda+\epsilon\kappa^2\left(G\frac{\partial {\cal L}_m}{\partial G}-{\cal L}_m\right)\right]\delta^\alpha_\beta + 2\epsilon\kappa^2 \frac{\partial {\cal L}_m}{\partial K} K^\alpha{}_\beta\right\}}}g^{\rho\nu}\,,
\ee
Accordingly, the parametrizations \eqref{RBG-Lag} and \eqref{GR-Lag} assume the following form
\be
\mathcal{L}_m(K,G) = \frac{\lambda}{\epsilon\kappa^2} - \left.\frac{1-\epsilon\kappa^2\left(2\frac{\partial\tilde{\cal L}_m}{\partial \tilde K}\tilde K + 2\frac{\partial\tilde{\cal L}_m}{\partial \tilde G}\,\tilde G - \tilde{\cal L}_m\right)}{\epsilon\kappa^2\sqrt{\det{\left\{\left[1 - \epsilon\kappa^2 \left(\frac{\partial\tilde{\cal L}_m}{\partial \tilde G}\,\tilde G - \tilde{\cal L}_m\right)\right]\delta^\rho_\sigma - 2 \epsilon\kappa^2\frac{\partial\tilde{\cal L}_m}{\partial \tilde K}\tilde K^\rho{}_\sigma \right\}}}}\right|_{\tilde K(K,G),\tilde G(K,G)}\,,\label{Lm}
\ee
\be
\tilde{\cal L}_m(\tilde K,\tilde G) = \left.\frac{\lambda + \epsilon\kappa^2 \left( 2\frac{\partial{\cal L}_m}{\partial K}K + 2\frac{\partial{\cal L}_m}{\partial G}\,G -\mathcal{L}_m \right)}{\epsilon\kappa^2\sqrt{\det{\left\{\left[\lambda + \epsilon\kappa^2 \left(\frac{\partial{\cal L}_m}{\partial G}\,G -\mathcal{L}_m(g,\psi)\right)\right]\delta^\rho_\sigma + 2\epsilon\kappa^2\frac{\partial{\cal L}_m}{\partial K}K^\rho{}_\sigma\right\}}}}\right|_{K(\tilde K,\tilde G),G(\tilde K,\tilde G)} - \frac{1}{\epsilon\kappa^2}\,.\label{TildeLm}
\ee

Using the multiplication rule \eqref{KK}, the mapping between metrics given by \eqref{EiBI-g-q} can be rephrased as the following relations between invariants
\bea
K&=&\frac12 g^{\mu\nu}F_{\mu\rho}g^{\rho\sigma}F_{\rho\nu} = \Omega_{GR}\left\{\tilde K - 2\epsilon\kappa^2\left[2\left(\tilde K^2+2\tilde G^2\right)\frac{\partial\tilde{\cal L}_m}{\partial \tilde K} + \tilde K\tilde G \frac{\partial\tilde{\cal L}_m}{\partial \tilde G} - \tilde K \tilde{\cal L}_m\right] + \right.\\
&+&\left.\epsilon^2\kappa^4\left[\left(\tilde{\cal L}_m - \tilde G \frac{\partial\tilde{\cal L}_m}{\partial \tilde G}\right)\left(\tilde K\tilde{\cal L}_m -4\tilde K^2 \frac{\partial\tilde{\cal L}_m}{\partial \tilde K} -8 \tilde G^2 \frac{\partial\tilde{\cal L}_m}{\partial \tilde K}-\tilde K \tilde G \frac{\partial\tilde{\cal L}_m}{\partial \tilde G}\right) + 4\tilde K\left(\tilde K^2 + 3\tilde G^2\right)\left(\frac{\partial\tilde{\cal L}_m}{\partial \tilde K}\right)^2\right]\right\}\,,\label{KtilKTilG}
\eea
\be
G=\Omega_{GR}^{1/2}\tilde G\,,\label{GtilKTilG}
\ee
where
\be
\Omega_{GR} = \frac{1}{\det\left\{\left[1 + \epsilon\kappa^2\left(\tilde{\cal L}_m - \tilde G \frac{\partial\tilde{\cal L}_m}{\partial\tilde G}  \right)\right]\delta^\alpha_\beta - 2\epsilon\kappa^2 \frac{\partial\tilde{\cal L}_m}{\partial\tilde K} \tilde K^\alpha{}_\beta \right\}}\,.
\ee
Given a metric that is a solution of GR coupled to matter fields through $\tilde{\cal L}_m$, one can generate a corresponding solution of the EiBI gravity using  \eqref{EiBI-g-q} while the matter sector conterpart is given by \eqref{Lm} where $\tilde K(K,G)$ and $\tilde G(K,G)$ comes from the inversion of the  mappings \eqref{KtilKTilG} and \eqref{GtilKTilG}. This is basically the procedure to follow to find a solution of the EiBI gravity associated to a known soltion of GR.

Viceversa, if a solution $g_{\mu\nu}$ of the EiBI gravity with a given matter sector ${\cal L}_m$ is known, than the GR counterpart $q_{\mu\nu}$ and $\tilde{\cal L}_m$ can be found from \eqref{EiBI-q-g} and \eqref{TildeLm} respectively, with $\tilde K(K,G)$ and $\tilde G(K,G)$ deduced from the following mapping between invariants based on \eqref{EiBI-q-g}
\bea
\tilde K &=&\frac12 q^{\mu\nu}F_{\mu\rho}q^{\rho\sigma}F_{\rho\nu} = \Omega^{-1}_{EiBI}\left\{\lambda^2  K + 2\epsilon\kappa^2\lambda\left[2\left( K^2+2 G^2\right)\frac{\partial{\cal L}_m}{\partial  K} +  K\,G\, \frac{\partial{\cal L}_m}{\partial  G} -  K {\cal L}_m\right] + \right.\\
&+&\left.\epsilon^2\kappa^4\left[\left({\cal L}_m -  G \frac{\partial{\cal L}_m}{\partial  G}\right)\left(K{\cal L}_m -4 K^2 \frac{\partial{\cal L}_m}{\partial  K} -8  G^2 \frac{\partial{\cal L}_m}{\partial  K}- K G \frac{\partial{\cal L}_m}{\partial  G}\right) + 4 K\left( K^2 + 3 G^2\right)\left(\frac{\partial{\cal L}_m}{\partial  K}\right)^2\right]\right\}\,,\label{TilKKG}
\eea
\be
\tilde G=\Omega_{EiBI}^{-1/2} G\,,\label{TilGKG}
\ee
where
\be
\Omega_{EiBI} = \det\left\{\left[\lambda - \epsilon\kappa^2\left({\cal L}_m - G \frac{\partial{\cal L}_m}{\partial G}  \right)\right]\delta^\alpha_\beta + 2\epsilon\kappa^2 \frac{\partial{\cal L}_m}{\partial K} K^\alpha{}_\beta \right\}\,.
\ee
%It is interesting to note that the has the same functional form when $\lambda = 1$ up to a sign flipping acting on $\epsilon$. 

%%%%%%%%%%%%%%%%%%%%%%%%%%%%%%%%%%%%%%%%%%%%%%%%%%%%%%%%%%%%%%%%%
\subsection{Examples}
%%%%%%%%%%%%%%%%%%%%%%%%%%%%%%%%%%%%%%%%%%%%%%%%%%%%%%%%%%%%%%%%%

The application of the generating technique can be made more concrete when the matter sector is specified. The simplest instance is the pure EiBI theory, namely the case where ${\cal L}_m = 0$. The corresponding matter sector in the GR frame can be found using \eqref{GR-Lag}, resulting in just a cosmological constant that depends on the parameters $\epsilon$ and $\lambda$ as follows
\be
\tilde{\cal L}_m = \frac{\lambda-1}{\epsilon\kappa^2\lambda}\,.
\ee
This means that any vacuum solution of EiBI gravity can be reinterpreted as an expanding or contracting solution in GR depending on the values of the parameters $\epsilon$ and $\lambda$. In particular the metrics in different frames are related by the following simple conformal transformation modulated by the parameter $\lambda$ :
\be
q_{\mu\nu} = \lambda g_{\mu\nu}
\ee
putting forward the possibility of interpreting cosmological expansion or contraction as a vacuum effect of the EiBI gravity when one shifts to the GR frame. Notice that this result differs from the usual expansion in power of $\epsilon$ of the EiBI theory that leads to GR plus a divergent cosmological constant when $\epsilon$ approaches zero \cite{Bazeia:2015uia}.

\vspace{0.5cm}

Another simple application is the case of a RBG described by a free Maxwell field coupled to EiBI gravity where the matter Lagrangian is represented by the usual kinetic term of the electromagnetism ${\cal L}_m=K/2$. According to \cite{Delhom:2019zrb}, the matter counterpart in the GR frame turns out to be 
\be
\tilde{\cal L}_m = \frac{1-2\lambda\pm \sqrt{1+ 2\epsilon\kappa^2\lambda\left(\tilde K-2\epsilon\kappa^2\lambda \tilde G^2\right)}}{2\epsilon\kappa^2\lambda}\, .
\ee
that reduces to BI electrodynamics \cite{Born:1934gh}
\be
\tilde {\cal L}_{BI} = \beta^2\left(1-\sqrt{1
-\frac{\tilde K}{\beta^2} - \frac{\tilde G^2}{\beta^4}}\right)= \beta^2\left(1-\sqrt{1
+\frac{1}{2\beta^2}F_{\mu\nu}\tilde{F}^{\mu\nu} - \frac{1}{16\beta^4}(F_{\mu\nu}{}^\star \tilde{F}^{\mu\nu})^2}\right)\ ,\label{BI-electr}
\ee
with $\beta=-1/(2\epsilon\kappa^2)$, when $\lambda = 1$ and the positive sign is chosen. This result is an application of the direct correspondence obtained from \eqref{TildeLm} after performing the substitutions
\be
K = \frac{2\left(\tilde K - 4\epsilon\kappa^2 \lambda\tilde G^2\right)\left(1+\epsilon\kappa^2\lambda\tilde K \pm \sqrt{1+ 2\epsilon\kappa^2\lambda\tilde K-4\epsilon^2\kappa^4 \lambda^2 \tilde G^2}\right)}{\epsilon^2\kappa^4\left(4\tilde G^2 + \tilde K^2\right)}\nonumber
\ee
\be
G = -\frac{2\tilde G\left[1-4\epsilon\kappa^2\lambda\tilde K + 2\epsilon^2\kappa^4 \lambda^2 \tilde G^2   \pm \left(1+\epsilon\kappa^2\lambda\tilde K\right) \sqrt{1+ 2\epsilon\kappa^2\lambda\tilde K-4\epsilon^2\kappa^4 \lambda^2 \tilde G^2}\right]}{\epsilon^2\kappa^4\left(4\tilde G^2 + \tilde K^2\right)}\nonumber \ .
\ee
that are obtained from the inversion of the relations \eqref{KtilKTilG} and \eqref{GtilKTilG}. Furthermore, the mapping between the metrics can be read from \eqref{EiBI-g-q} and \eqref{EiBI-q-g} leading to
\be
g_{\mu\nu} = \pm\frac{\left[1\pm\sqrt{1+ 2\epsilon\kappa^2\lambda\left(\tilde K-2\epsilon\kappa^2\lambda \tilde G^2\right)}\right]q_{\mu\nu} + 2\epsilon\kappa^2\lambda\tilde K_{\mu\nu}}{2\lambda \sqrt{1+ 2\epsilon\kappa^2\lambda\left(\tilde K-2\epsilon\kappa^2\lambda \tilde G^2\right)}}\,,\label{EiBI-Maxw-g-q}
\ee
\be
q_{\mu\nu}=\frac12\left(2\lambda+\epsilon\kappa^2 K\right)g_{\mu\nu}-\epsilon\kappa^2 K_{\mu\nu}\,.\label{EiBI-Maxw-q-g}
\ee
These explicit maps between metrics complement the correspondence discovered in \cite{Delhom:2019zrb} providing a very useful recipe to generate solutions of the EiBI theory coupled to Maxwell electromagnetism from solutions of GR coupled to BI electrodynamics and viceversa.

\vspace{0.5cm}

The simplest theory that brings together GR and electromagnetism has just a free Maxwell field in its matter sector, that is equivalent to say that $\tilde{\cal L}_m=\tilde K/2$. Due to the large literature on the solutions of this kind of theory, it is of great relevance to apply the method described in the previous sections to find the corresponding EiBI theory for this case. Contrary to the previous examples, the matter Lagrangian in the RBG frame is unkown. According to the procedure, the first step consists in specifying the relations between invariants \eqref{KtilKTilG} and \eqref{KtilKTilG} by a direct substitution of $\tilde{\cal L}_m$ that leads to
\be
K=4\frac{4\tilde K -\epsilon\kappa^2\left(4-\epsilon\kappa^2\tilde K\right)\left(\tilde K^2 + 4\tilde G^2\right)}{4-\epsilon^2\kappa^4\left(\tilde K^2 + 4\tilde G^2\right)}\,,\qquad G=\frac{4\tilde G}{4-\epsilon^2\kappa^4\left(\tilde K^2 + 4\tilde G^2\right)}\,.
\ee
From the inversion of these coupled equations, one obtains the following substitutions :
\be
\tilde K = \frac{2\left(K+4\epsilon\kappa^2G^2\right)\left[1-\epsilon\kappa^2 K\mp\sqrt{1-2\epsilon\kappa^2 \left(K + 2\epsilon\kappa^2 G^2\right)}\right]}{\epsilon^2\kappa^4\left(K^2+4G^2\right)}
\ee 
\be
\tilde G = -\frac{2G\sqrt{1-2\epsilon\kappa^2 \left(K + 2\epsilon\kappa^2 G^2\right)} \left[\sqrt{1-2\epsilon\kappa^2 \left(K + 2\epsilon\kappa^2 G^2\right)}\mp \left(1-\epsilon\kappa^2 K\right)\right]}{\epsilon^2\kappa^4\left(K^2+4G^2\right)}
\ee 
that are the fundamental ingredients to unveil the following matter interactions in the RBG frame using \eqref{Lm} and \eqref{TildeLm} :
\be
{\cal L}_m = \frac{2\lambda-1\mp\sqrt{1-2\epsilon\kappa^2 \left(K + 2\epsilon\kappa^2 G^2\right)}}{2\epsilon\kappa^2}\,.
\ee
Notice that the particular choice $\lambda=1$ reduces this Lagrangian to the one describing the BI electrodynamics \eqref{BI-electr} when $\beta = \frac{1}{\epsilon\kappa^2}$ and the minus sign is choosen. Surprisingly, the interplay between the nonlinearities in the gravity sector and the matter sector play a fundamental role to cancel the nonlinear interactions in the action when going from the RBG frame to GR frame. Concerning  the mappings between metrics, after some algebra the mappings \eqref{EiBI-g-q} and \eqref{EiBI-q-g} specified to the matter Lagrangians just found, leads to the following relations
\be
g_{\mu\nu}=\frac12\left(2-\epsilon\kappa^2\tilde K\right)q_{\mu\nu} + \epsilon\kappa^2\tilde K_{\mu\nu}\,,
\ee
\be
q_{\mu\nu} = \frac{\left[\sqrt{1-2\epsilon\kappa^2 \left(K + 2\epsilon\kappa^2 G^2\right)}\pm1\right]g_{\mu\nu} \mp 2\epsilon\kappa^2 K_{\mu\nu}}{2\sqrt{1-2\epsilon\kappa^2 \left(K + 2\epsilon\kappa^2 G^2\right)}}\,.
\ee
These three concrete examples explain explicitly how to apply the machinery of Sec.\eqref{GenSol} to all possible scenarios that can arise when seeking for a classical equivalence between EiBI and GR that involves electromagnetic interactions. It is worth to emphasize that the possible applications are much broader since any electrodynamic can be choosen as a matter sector of any frame to settle a correspondence using \eqref{Lm} and \eqref{TildeLm} that can be used to find solutions of the EiBI gravity.

%%%%%%%%%%%%%%%%%%%%%%%%%%%%%%%%%%%%%%%%%%%%%%%%%%%%%%%%%%%%%%%%%
\section{Conclusions}
%%%%%%%%%%%%%%%%%%%%%%%%%%%%%%%%%%%%%%%%%%%%%%%%%%%%%%%%%%%%%%%%%

In this contribution we refined in a self-contained way a general method to establish an on-shell classical correspondence between a wide family of modified gravities and GR. Special attention has been dedicated to the mapping between metrics in RBG and GR, providing previously unknown explicit equations ruling the relations between solutions in different frames for any RBG and matter sector. In order to present a concrete application of the procedure, the non-trivial case of EiBI gravity has been worked out in full detail, providing a general recipe to discover a correspondence for any matter content. The specific case of nonlinear theories involving electromagnetic field minimally coupled to EiBI, served as a toy model to further explain the application of the general method and to provide explicit example for the simplest cases of matter interactions. In particular, when matter fields are absent, the EiBI theory can be interpreted as GR with cosmological constant in a different frame.
The major result in this work is a general strategy to produce new exact solutions for RBG starting from known solutions of GR setting the stage for an extensive study of the physics behind modified gravity theories. It is expected that extentions of the analysis to matter sources of interest for observational and theoretical applications beyond scalar and electromagnetic fields, will be a straightforward task using the general framework developed in the present paper. This will give access to previously unknown astrophysical and cosmological scenarios overcoming the complexity of solving the field equations associated to RBGs.

%%%%%%%%%%%%%%%%%%%%%%%%%%%%%%%%%%%%%%%%%%%%%%%%%%%%%%%%%%%%%%%%%
\section*{Acknowledgments}
%%%%%%%%%%%%%%%%%%%%%%%%%%%%%%%%%%%%%%%%%%%%%%%%%%%%%%%%%%%%%%%%%

The author thanks Prof. Luis C.B. Crispino for the kind invitation to take part to the ``V Amazonian Symposium on Physics"  in Bel\'em (Brazil), where part of this work has been elaborated.

%%%%%%%%%%%%%%%%%%%%%%%%%%%%%%%%%%%%%%%%%%%%%%%%%%%%%%%%%%%%%%%%%

\end{document}